\documentclass[prc,twocolumn]{revtex4}
\usepackage{amsmath,amssymb,epsfig,bm}
\newcommand{\be}{\begin{equation}}
\newcommand{\ee}{\end{equation}}
\newcommand{\bea}{\begin{eqnarray}}
\newcommand{\eea}{\end{eqnarray}}

\begin{document}
\title{Gravitational radiation from crystalline color-superconducting hybrid 
stars}

\author{Bettina~Knippel, Armen~Sedrakian}
\address{Institute for Theoretical Physics,
Goethe-University, D-60438 Frankfurt am Main, Germany}

\date{\today}

\begin{abstract}
The interiors of high mass compact (neutron) stars may contain deconfined
quark matter in a crystalline color superconducting (CCS) state. On a
basis of microscopic nuclear and quark matter equations of states we explore
the internal structure of such stars in general relativity. We find that 
their stable sequence harbors CCS quark cores with masses 
$M_{\rm core}\le (0.78-0.82)M_{\odot}$ and radii $R_{\rm core}\le 7$ km. 
The CCS quark matter can support nonaxisymmetric deformations, because 
of its finite shear modulus, and can generate 
gravitational radiation at twice the rotation 
frequency of the star. Assuming that the CCS core is maximally
strained we compute the maximal quadrupole moment it can sustain.
The characteristic strain of gravitational wave emission $h_0$ predicted
by our models are compared to the upper limits obtained by the LIGO and GEO 600
detectors. The upper limits are consistent with the breaking strain of 
 CCS matter $\sigma \le 10^{-4}$ and large pairing gaps 
$\Delta\sim 50$ MeV,
or, alternatively, with $\sigma \sim 10^{-3}$ and small pairing gaps
$\Delta\sim 15$ MeV. An observationally determined value of the characteristic 
strain $h_0$ can pin down  the product
$\sigma\Delta^2$. On the theoretical side a better understanding of the
breaking strain of CCS matter will be needed to predict reliably the level
of the deformation of CCS quark core from first principles.

\end{abstract}


\maketitle

\section{Introduction}
\label{sec:intro}

Compact (neutron) stars 
are important sources of gravitational wave radiation. 
Gravity waves, to lowest order, arise from time-dependent 
quadrupole deformations of masses.  Rotating, isolated 
compact stars will emit gravitational radiation if their 
mass distribution is nonaxisymmetric with respect to their 
rotation axis. The axial symmetry can be broken over detectable 
time-scales in a number of ways, e. g.,  through distortions 
in the solid phases of the star~\cite{DIST1,DIST2,DIST3,Cutler:2002np,
Haskell:2006sv,Chamel:2008ca,Ushomirsky:2000ax}, 
deformations caused by strong magnetic 
fields~\cite{Bonazzola:1995rb,Cutler:2002nw,Wasserman:2003,
Akgun:2007ph,Haskell:2007bh}, or by 
precession~\cite{PR1,PR2,Sedrakian:1998vi,Cutler:2000bp,Jones:2001yg}. 
Continuous gravitational waves emitted by nonaxisymmetric 
rotating compact stars are expected to be in the 
bandwidth of current gravitational wave interferometric detectors. 
Upper limits on the strain of gravitational waves 
from a large selection of known radio pulsars were set 
recently by the LIGO Collaboration under the assumption that the 
radiation is at twice the pulsar spin frequency~\cite{Abbott:2007}. 
Ellipticities derived from these upper limits are within a range 
that is compatible with theoretical predictions.
Current data from the fifth LIGO science run places upper limits on the 
gravitational wave amplitude from the Crab pulsar 
at $h = 3.4\times 10^{-25}$ (95\% confidence), which is below 
the spin-down limit (i.e., the limit obtained assuming that all the 
energy loss is due to gravitational radiation) by a factor of 
4~\cite{LIGO_S5_CRAB}.

In this work we study gravitational wave emission from compact 
stars featuring a crystalline color-superconducting (CCS) phase 
of quark matter in their interiors. 
Our study is based on  microscopic equations 
of state of nuclear and quark matter, described earlier in a
study of the integral parameters of nonrotating and rapidly 
rotating hybrid configurations~\cite{Ippolito:2007hn}. 
The ellipticity of CCS quark stars and the associated gravitational wave 
emission was estimated earlier for  uniform-density incompressible 
models. On the basis of such a model Lin~\cite{Lin:2007} 
concluded that the upper limit for the Crab pulsar provides useful 
constraints on the QCD parameters. 
Haskell et al.~\cite{Haskell:2007} estimated core deformations and 
associated with them constraints on the QCD parameters for sequences of 
1.4 $M_{\odot}$ and $R=10$ km 
stars containing CCS quark cores with different transition densities 
from nuclear to quark matter. The nuclear matter equation of 
state was approximated by a $n=1$ polytrope, and quark matter  
was treated as an incompressible fluid. Equilibrium and stability of 
hybrid compact stars constructed from microscopic equations of state of quark and 
nuclear matter were studied in Ref.~\cite{Ippolito:2007hn}. 
For stellar sequences that are based on microscopic input the stability 
is not a guaranteed feature. 
The equilibrium among the quark and nuclear phases 
is achieved for rather stiff nuclear equations of state, 
while those configurations that contain quark phase(s) often 
belong to unstable branches of the sequences
~\cite{Buballa:2003et,Baldo:2002ju,Ma:2007iw,Pagliara:2007ph,
Grigorian:2003vi,Blaschke:2007ri,Alford:2004pf}. 
For our microscopic input a stable branch of CCS hybrid 
stars emerges in the form of 
a ``second family'' of configurations, which is separated from their 
purely nuclear counterparts by a region of instability. The masses of 
the resulting hybrid configurations are close to the maximum 
sustainable mass $M_{\rm max}\simeq 2M_{\odot}$, 
i.e., are much more massive than ``canonical'' neutron 
stars with $M\sim 1.4 M_{\odot}$. 

Below, by resorting to microphysical input we add realism to
the treatment of the gravitational radiation from hybrid compact 
stars, in particular, the constraints on microscopic parameters 
derived from the observational upper limits. 
Even though  we lift a large portion of 
uncertainties related to the microphysical input of the theory,
the elastic properties of the CCS phase and other solid phases 
of our models (nuclear crusts) remain uncertain.
The breaking strain that solid phases of compact stars can sustain 
is difficult to compute and is usually assumed to lie in the range 
$10^{-5}\le \sigma\le 10^{-2}$. The values close to the upper 
``optimistic'' bound of 
$\sigma$ implies distortions in the star's crusts and/or core that 
are interesting from the perspective of gravitational wave emission.
We concentrate below on distortions in the CCS phase and will ignore 
any other contributions, as those originating from the distortions 
in the crusts, precession or magnetic deformations. Distortions may exist
in other types of compact objects, such as solid strange stars, quark stars, 
stars with mixed phases of quarks, meson-condensates and 
hadrons~\cite{WEBER_BOOK,Sedrakian:2006mq,Brown:2008cn,Blaschke:2008cu}; 
the problem of gravitational wave emission 
from such objects is discussed in the literature~\cite{Xu:03,Owen:05}.
Furthermore, we assume that the CCS quark matter is the true ground 
state of quark matter in some density 
interval~\cite{Alford:2000ze,Bowers:2002xr,Kiriyama:2006ui,
Casalbuoni:2005zp,Ippolito:2007uz,Rajagopal:2006ig,
Mannarelli:2006fy,Mannarelli:2008zz} and that the 
value of the dynamical strange quark mass is small enough to favor 
the three-flavor variant of this phase~\cite{Casalbuoni:2005zp,
Mannarelli:2006fy,Ippolito:2007uz,Rajagopal:2006ig}. 
However, among the candidate phases for the ground state of quark 
matter at intermediate densities, the three-flavor CCS quark phase 
and its two-flavor counterpart are the only phases that behave as solids 
(nonzero shear modulus). Therefore, their possible manifestations via
gravitational wave emission is  unique among the 
color-superconducting phases.

This paper is organized as follows: In Sec.~\ref{sec:models} we 
study the equilibrium sequences of nonrotating hybrid compact stars 
featuring CCS quark cores. We compute the masses, radii, and 
quadrupole moments of the quark cores of hybrid configurations.
Section~\ref{sec:grav_rad} discusses the gravitational wave radiation 
from the CCS quark cores. The experimental upper limits on the strain 
of gravitational waves from pulsars are compared to the theoretical 
predictions of our models. Our conclusions are collected in 
Sec.~\ref{sec:conclusion}.

\section{Models of CCS quark stars}
\label{sec:models}

We consider models of hybrid compact stars constructed from zero-temperature
microscopic equations of state of nuclear matter described by 
Dirac-Bruckner-Hartree-Fock (DBHF) theory and quark matter described 
by the Nambu-Jona-Lasinio  model (see also Ref.~\cite{Ippolito:2007hn}).
The DBHF approach is based on a self-consistent solution of the equation 
for the in-medium relativistic $T$ matrix 
and nucleon self-energy starting from a 
realistic nucleon-nucleon potential. The potential is adjusted to reproduce
the nuclear phase shifts and deuteron binding energy. The saturation properties 
of nuclear matter are well reproduced within  DBHF theories. The quark equation 
of state is derived from the Lagrangian of the Nambu--Jona-Lasinio model. 
This model is based on the four-quark contact interaction picture, i.e., the 
gluons are integrated out from the theory. It exhibits the chiral 
symmetry restoration 
in quark matter, but lacks confinement. Its coupling constants and momentum
cutoff are adjusted to reproduce the meson masses and the pion decay constant
in the vacuum. Deconfinement phase transition 
between the nuclear and quark phases is implemented 
via a Maxwell construction, i.e., it occurs at the 
point where the pressures of both phases as functions of the baryonic 
chemical potential coincide. At the deconfinement phase transition the 
density experiences a jump at constant pressure and zero temperature.
Within the quark matter phase a Cooper pair condensate of quarks with  
mismatched Fermi-surfaces will emerge. Our models feature the so-called
three-flavor Larkin-Ovchinnikov-Fulde-Ferrell (LOFF) or 
crystalline color-superconducting (CCS) phase as a 
possible candidate of the ground state of Cooper-paired 
quark matter~\cite{Alford:2000ze,Bowers:2002xr,Mannarelli:2006fy,
Casalbuoni:2005zp,Ippolito:2007uz,Kiriyama:2006ui,Rajagopal:2006ig,
Mannarelli:2008zz}. 
In this phase the condensate order parameter is spatially modulated; as a 
result the superfluid phase behaves as a solid with finite shear modulus. 
This, in turn, implies that the LOFF phase can sustain equilibrium
nonaxisymmetrical deformations, which potentially can lead to gravitational 
wave radiation. It should be noted that a more complete treatment of 
high-density matter will need to include  phases other than the 
CCS phase and the possibility of transitions between 
them~\cite{Alford:2007xm}. However, the latter phases are not 
crystalline and are unlikely to affect the physics of gravitational 
wave emission in a direct manner. Indirectly, they may have an effect, 
e. g., by changing the fractional volume occupied by the CCS phase.

\begin{figure}[t]
\begin{center}
\includegraphics[height=8.0cm,width=\linewidth,angle=0]{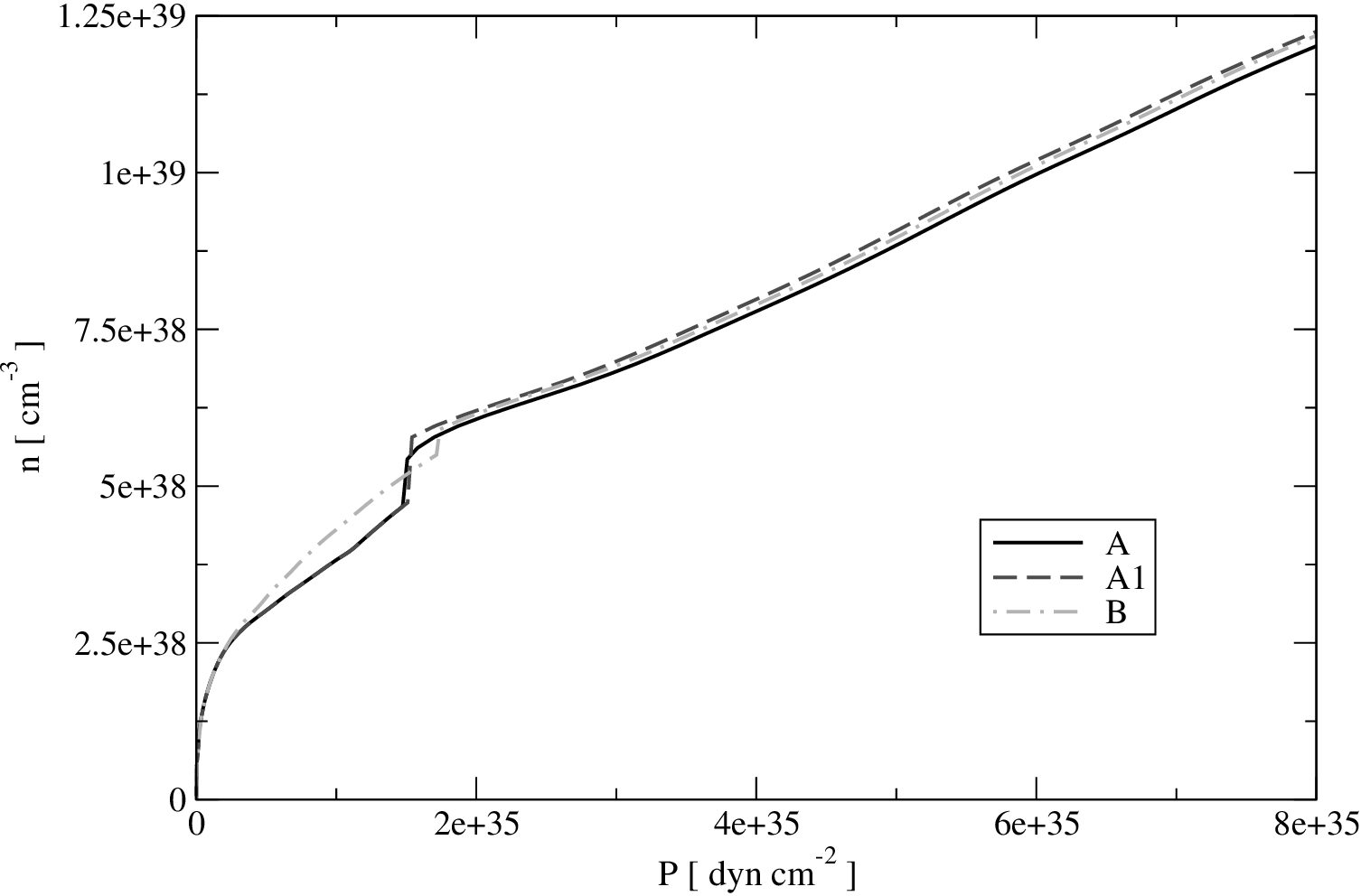}
\end{center}
\caption[]
{(Color online)
Number density versus pressure for models A ({\it solid, black online}),
A1 ({\it dashed, red online}), and B ({\it dashed-dotted, blue online}).
For  models A and A1 the  nuclear (low density) equation of
state is the same; for models A and B the  quark (high-density)
equation of state is the same. At  the deconfinement phase
transition there is a jump in the density at constant pressure.
}
\label{fig:n_P}
\end{figure}
\begin{figure}[tb]
\begin{center}
\includegraphics[height=8.0cm,width=\linewidth,angle=0]{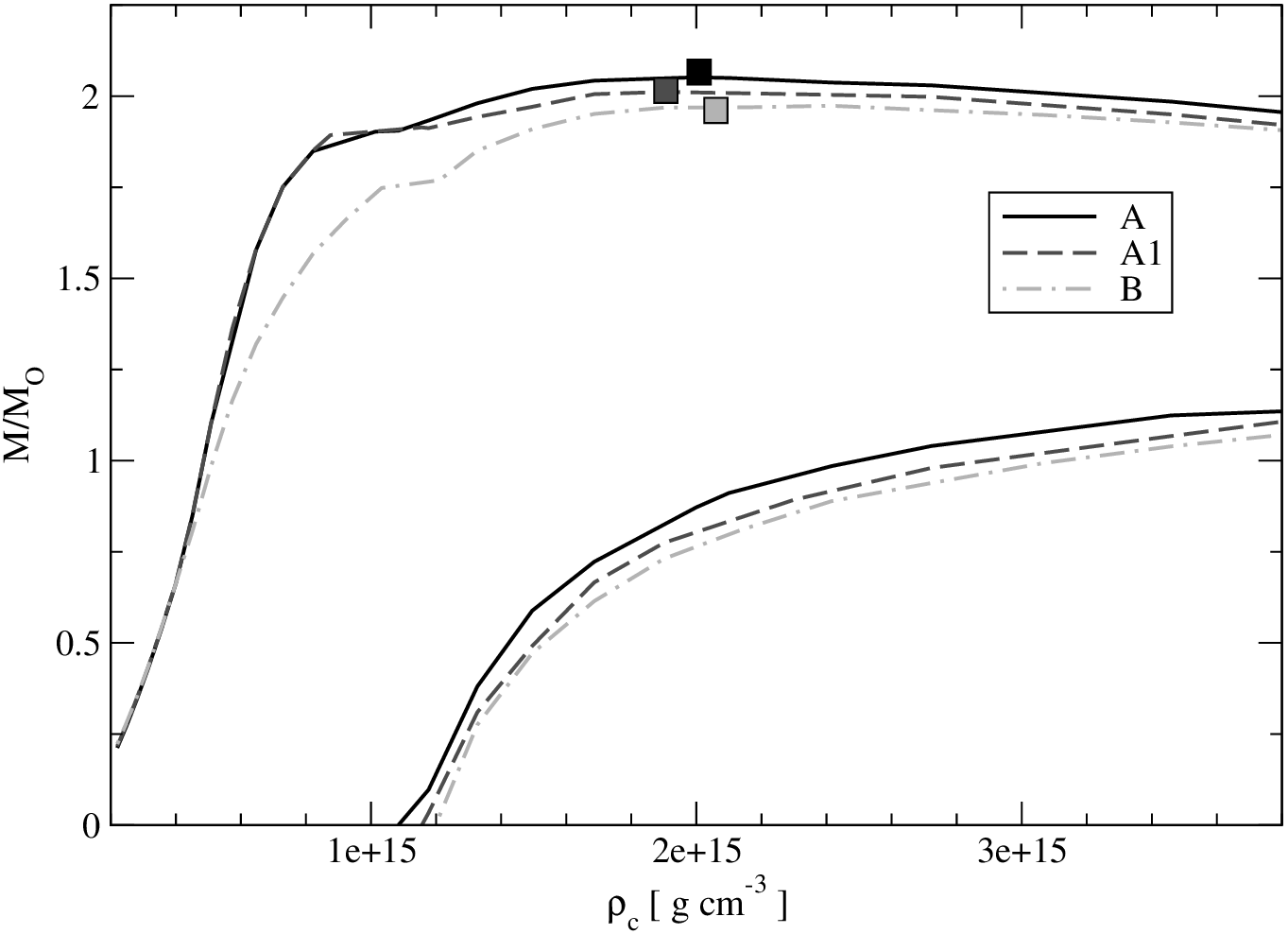}
\end{center}
\caption[]
{(Color online)
Dependence of the total stellar mass and the mass of the quark core
in units of solar mass $M_{\odot}$
on the central density for nonrotating configurations based on 
equations of state  A, A1, and B (the labeling 
is as in Fig.~\ref{fig:n_P}). The lower set of curves represents the 
masses of the CCS quark cores, the upper set - the total masses of the
configurations. The maximal masses are marked with boxes.
}
\label{fig:M_density}
\end{figure}

We consider three different equations of state that are displayed 
in Fig.~\ref{fig:n_P} (see also Fig.~1 of Ref.~\cite{Ippolito:2007hn}). 
For models A and A1 the  nuclear (low density) equation of
state is the same; for models A and B the  quark (high-density)
equation of state is the same. At  the deconfinement phase
transition there is a jump in the density at constant pressure. The 
high-density regime contains two equations of states for crystalline color 
superconductivity which differ by the normalization of pressure at zero 
density (bag constant). Model A1 is normalized such that the pressure 
vanishes at zero density. For models A and B the zero density pressure 
is shifted by amount $\delta p = 10$ MeV/fm$^3$. The quark condensate is 
assumed to be in the strong coupling regime, where 
the ratio of the couplings in the quark-quark and quark-antiquark channels
$\eta = 1$ (the ``canonical'' value is $\eta=0.75$ and follows from the 
Fierz transformation from the quark--anti-quark to quark-quark channel.) 
The values of the (baryonic) chemical potential (in MeV) and pressure 
(in MeV/fm$^3$) at the quark-nuclear matter interface are (1234.2; 96) 
for  model A1, (1230.0; 95) for   model A and (1234.8; 108) 
for  model B~\cite{Ippolito:2007hn}.  
\begin{figure}[tb]
\begin{center}
\includegraphics[height=8.0cm,width=\linewidth,angle=0]{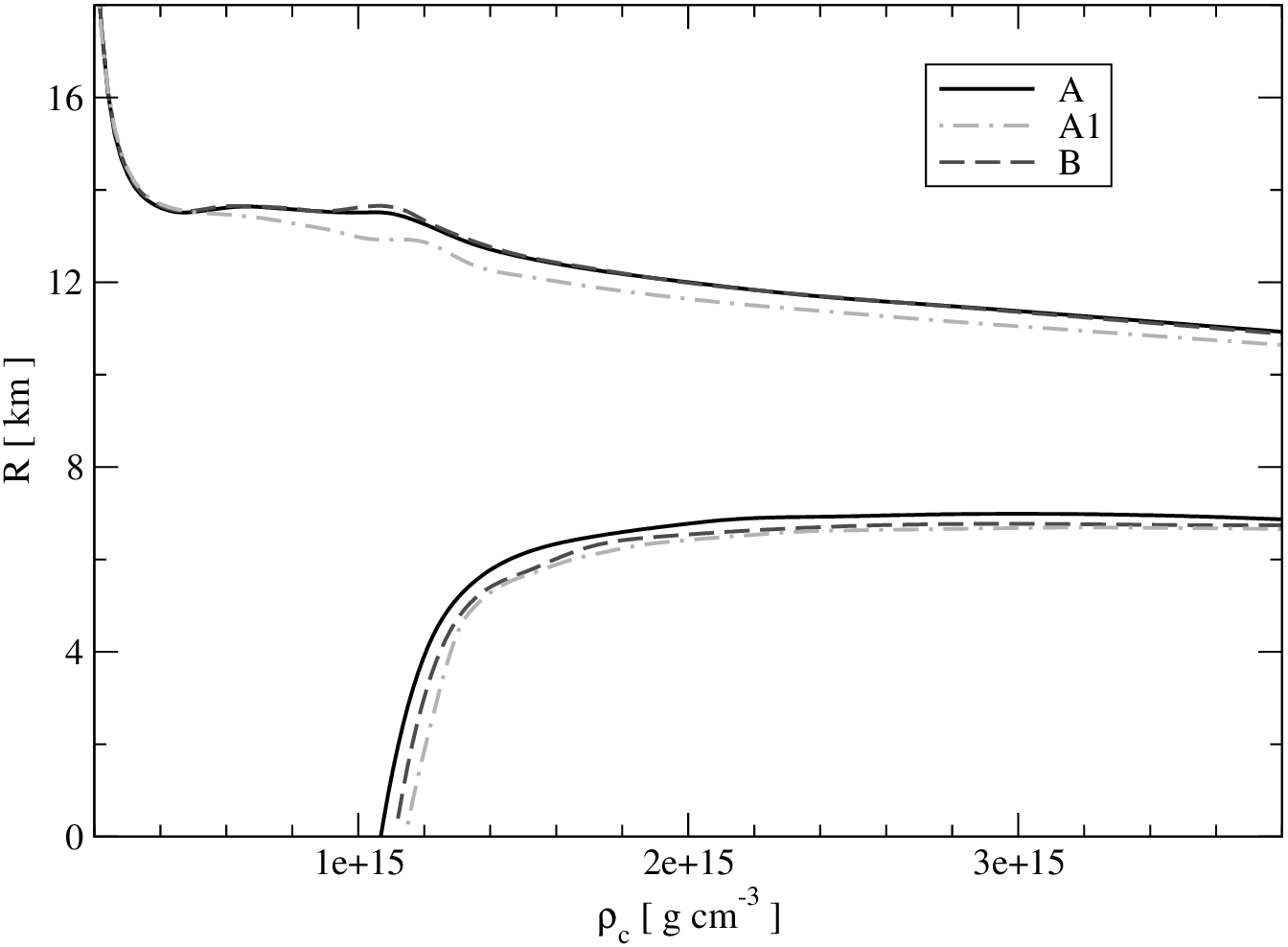}
\end{center}
\caption[]
{(Color online)
Dependence of radii of  nonrotating compact stars and the radii of their 
quark cores on  central density for models A, A1, and B
(the labeling is the same as in  Fig.~\ref{fig:n_P}). The lower set 
of curves represents the radii of the CCS quark cores, the upper set 
the radii of the configurations.
}\label{fig:r_cdens}
\end{figure}
We start by considering  the equilibrium and stability of 
nonrotating cold hybrid stars with crystalline color-superconducting 
cores. Our emphasis will be on the internal structure and the integral 
parameters (mass, radius, etc) of the quark cores, rather than the 
integral parameters of the stars (these are discussed in
Ref.~\cite{Ippolito:2007hn}).  We parameterize the sequences 
of equilibrium, nonrotating stellar configurations in general relativity 
in terms of the central density $\rho_c$ of the configuration. It is
assumed that the configurations are cold, i.e., the stellar structure 
does not depend on temperature. (Because of large magnitudes of gaps,
color-superconducting phases can affect the early thermal evolution 
of neutron stars; we assume that our models have cooled down to 
temperatures well below the respective Fermi 
energies~\cite{Schaab:1996gd,Grigorian:2004jq,Page:2005fq}). 
We obtain the spherically symmetric 
solutions of Einstein's equations for self-gravitating fluids by 
solving the well-known Tolman-Oppenheimer-Volkoff equations~\cite{TOV}. 
A sequence of configurations is stable if the derivative $dM/d\rho_c$ 
is positive, i.e.,  the mass is an increasing function of the 
central density. At the point of instability the fundamental (pulsation) 
modes become unstable. If the stability is regained at higher central 
densities the modes by which the stars become unstable toward the 
collapse belong to higher order harmonics.

The masses of the stellar configurations ($M$) and the CCS quark cores 
($M_{\rm core}$) as a function of the central density are shown 
in Fig.~\ref{fig:M_density}. The sequences of hybrid configurations 
($M_{\rm core} >  0)$ appear with increasing central density when the 
density of deconfinement phase transition is reached. Prior to that 
point the stars are purely nuclear and become unstable ($dM/d\rho_c\le 0$)
before the stability is regained by the hybrid configurations.
The masses of this ``second family'' vary in a narrow range around 
the maxim attainable mass  $2M_{\odot}$, which is reached at the central 
density $2 \times 10^{15}$ g cm$^{-3}$. The values of the maximum masses are 
large as a consequence of the hardness of the underlying nuclear 
equations of state. 
The masses of the CCS quark cores cover the range  $0\le M_{\rm core}
/M_{\odot}\le 0.75-0.88$ for central densities $1.3 \times 
10^{15}\le \rho_c\le 2\times 10^{15}$. Thus, the quark core mass ranges 
from one third to about the half of the total stellar mass. 
Above one solar mass quark cores can be harbored  
only by unstable configurations.

Fig.~\ref{fig:r_cdens} displays the radii of hybrid configurations 
along with the radii of the CCS quark cores. With the onset of CCS quark 
phase the radii of the configurations become smaller, i.e.,  the hybrid 
configurations are  more compact (and more massive) than their neighboring 
purely nuclear counterparts.  Within their stability range the hybrid 
configurations have almost constant radii $R=12$ km. The radius of the 
quark core increases rapidly with the onset of the quark phase and 
saturates at the asymptotic value $R_{\rm core} =7$ km.  The quark 
cores occupy about 20\% of the volume of the star.
\begin{figure}[tb]
\begin{center}
\includegraphics[height=8.0cm,width=\linewidth,angle=0]{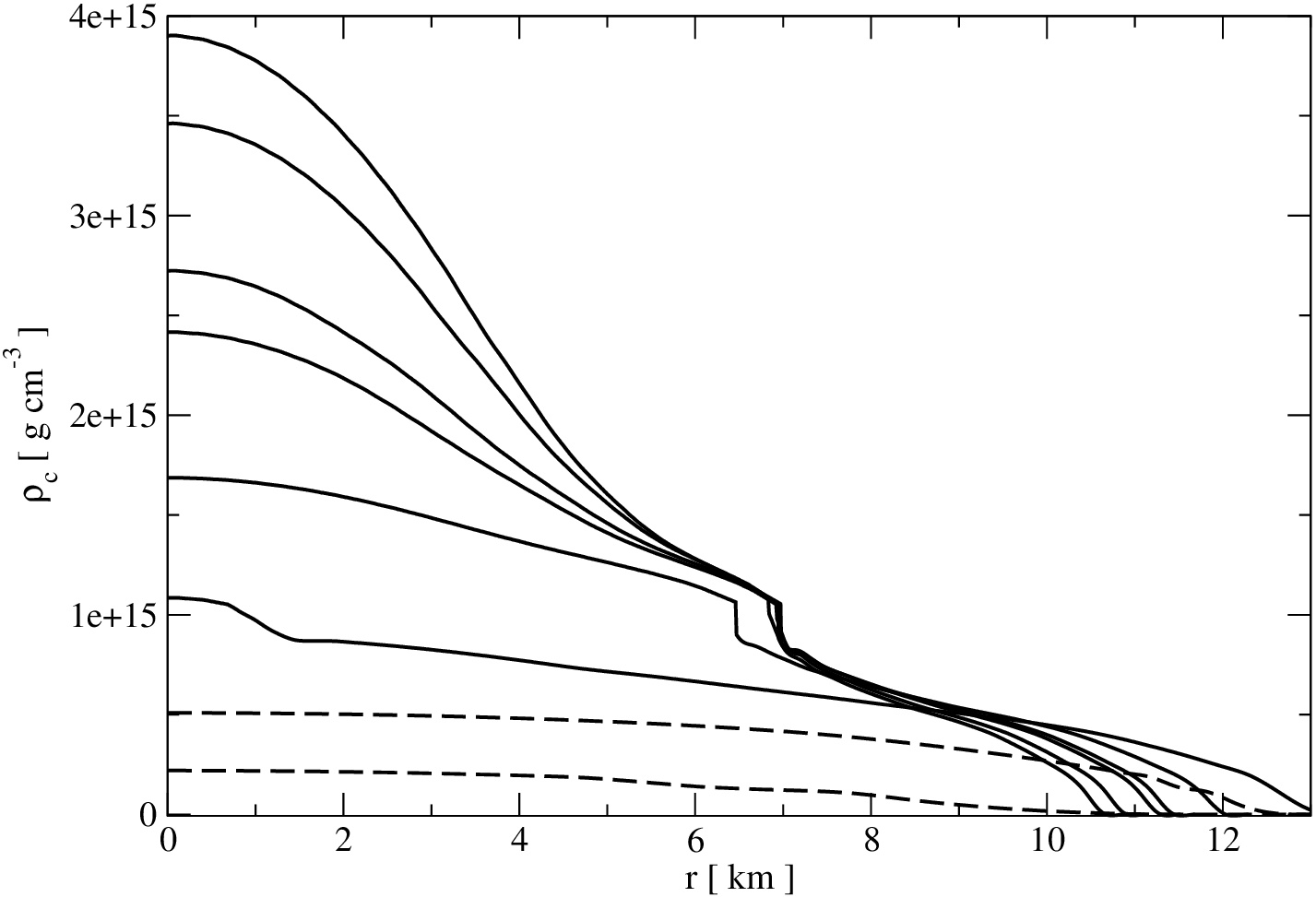}
\end{center}
\caption[]
{
Dependence of density on the internal radius for model A. 
The dashed lines correspond to purely nuclear stars, the solid 
lines  to hybrid configurations. For latter models there is a
density jump at the phase transition to quark matter. The transition 
radius saturates at the value $R_{\rm core}=7$ km as the central 
density is increased. 
}
\label{fig:density_profile}
\end{figure}
The density profiles of configurations are displayed in 
Fig.~\ref{fig:density_profile}. For purely nuclear configurations 
the density profile is flat in the core. As the central density 
exceeds the deconfinement transition density, the profiles show
a density jump at some internal radius. The jump reflects 
the behavior of the equation of state at the deconfinement phase 
transition.  There is a rapid growth of the radius (volume) 
of the quark phase as the density is increased from $1.3 
\times 10^{15}$ to $1.6 \times 10^{15}$ g cm$^{-3}$. For larger central 
densities, i.e.,  more massive objects, the radius at which the transition 
from quark to nuclear matter takes place is independent of the 
central density of the configuration (see also Fig.~\ref{fig:r_cdens}).
\begin{figure}[bth]
\begin{center}
\includegraphics[height=8.0cm,width=\linewidth,angle=0]{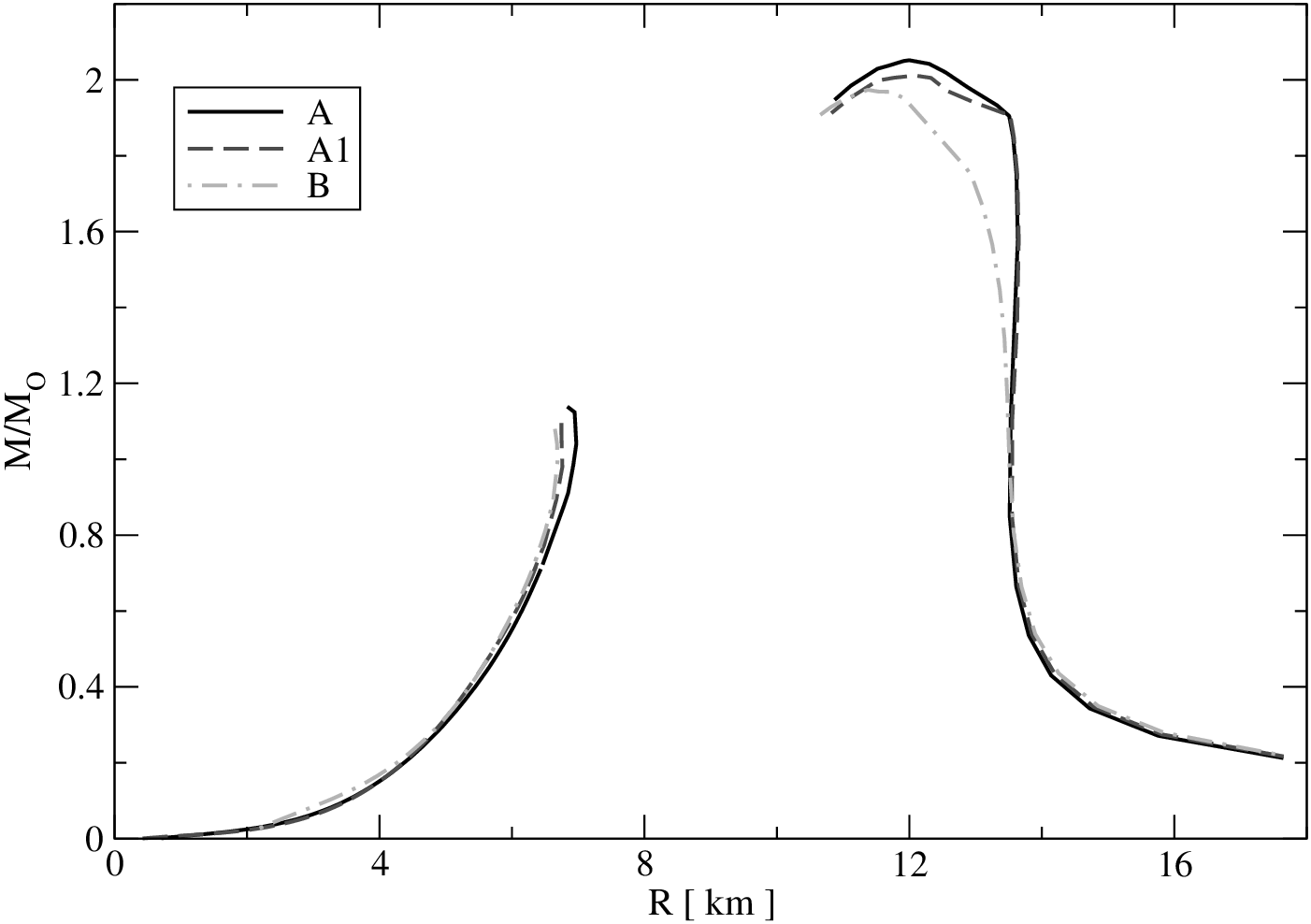}
\end{center}
\caption[]
{(Color online)
Mass-radius relationship for hybrid stars for  models A, A1, and B 
(the labeling is as in Fig.~\ref{fig:n_P}). The right set corresponds to 
the stellar configurations, the left set to the CCS quark cores. 
}
\label{fig:M_R}
\end{figure}

Figure ~\ref{fig:M_R} displays the mass-radius ($M$-$R$) relationship for 
the stellar configurations (right set of curves) and the CCS quark core
(the left set of curves) for models A, A1, and B. The $M$-$R$ diagram for
the stellar configurations has been discussed elsewhere~\cite{Ippolito:2007hn}; 
we recall that the
obtained mass-radius relationships are consistent with the current astronomical 
bounds on neutron star masses. They are consistent with the line in the $M$-$R$ 
diagram inferred from EXO0748-676~\cite{Ozel:2006bv}. Note that the
emission of redshifted lines used to put constraints in Ref.~\cite{Ozel:2006bv}  
has not been confirmed in a subsequent observation of Ref.~\cite{Cottam}. 
Also note that alternative models of quark stars are consistent with 
these bounds, which therefore do not rule out the hypothesis of quark matter 
in compact stars~\cite{Alford:2006vz}.  The  $M$-$R$ tracks for the CCS quark core 
display the rapid increase in the radius of the core in the presence of 
the small amount of quark matter [the $M(R)$ slope is almost horizontal 
to the $R$-axis]. The maximal masses of quark cores of the stable configurations 
are in the range $M\le 0.75-0.88 M_{\odot}$ depending on the equation of state; 
cores with larger masses correspond to configurations that are unstable
toward collapse into a black hole.

\section{Quadrupole moments and gravitational wave emission }
\label{sec:grav_rad}
The characteristic strain amplitude of gravitational waves 
emitted by a triaxial star rotating about its principal axis is  
\be \label{eq:h0}
h_0=\frac{16\pi^2 G}{c^4}\frac{\epsilon I_{zz}\nu^2}{r},
\ee
where $\nu$ is the star's rotation frequency, $r$ is the distance to the 
observer, $\epsilon=(I_{xx}-I_{yy})/I_{zz}$ is the equatorial ellipticity, 
$I_{ij}$ is the tensor of moment of inertia, $G$ is the
gravitational constant, $c$ is the speed of light. The characteristic
amplitude, Eq. (\ref{eq:h0}), does not involve the orientation of the 
source with respect to the observer; it is related to the sources' 
angle averaged field strength  $\langle h^2 \rangle 
= \int {d\Omega}(h_{+}^2+h_{x}^2)/4\pi $ by the relation 
$h_0 \simeq 1.15 \langle h^2 \rangle$. Alternatively, the 
strain amplitude can be expressed in terms of the $m=2$ mass 
quadrupole moment as 
\be \label{eq:h0_Q22}
h_0=\frac{16\pi^2 G}{c^4}\left(\frac{32\pi}{15}\right)^{1/2}
\frac{ Q_{22}\nu^2}{r},
\ee
where the mass multipole is defined as $Q_{lm}=\int \rho r^lY_{lm}^* d^3r$.
Thus, to compute the characteristic strain amplitude of gravitational waves
we need to compute the quadrupole moment of a CCS quark core in static 
equilibrium. The elastic deformations are assumed to be small perturbation 
on the background equilibrium of the star; the equilibrium between gravity 
and elastic forces, in the Cowling approximation, implies 
that~\cite{Ushomirsky:2000ax}
\be\label{eq:perturbation} 
\nabla^i\delta \tau_{ij} =\delta \rho g(r) \hat r_j,
\ee
where $g(r)$ is the local gravitational acceleration, 
$\delta$ represents Eulerian perturbation, 
$\tau_{ij}=-p g_{ij}+ t_{ij}$ is the stress-energy 
tensor of the CCS phase, $t_{ij}$ is the shear stress tensor,
and $g_{ij}$ is the flat 3-metric (indices $i$ and $j$ run over 
1,2,3). 
\begin{figure}[t]
\begin{center}
\includegraphics[height=8.cm,width=\linewidth,angle=0]{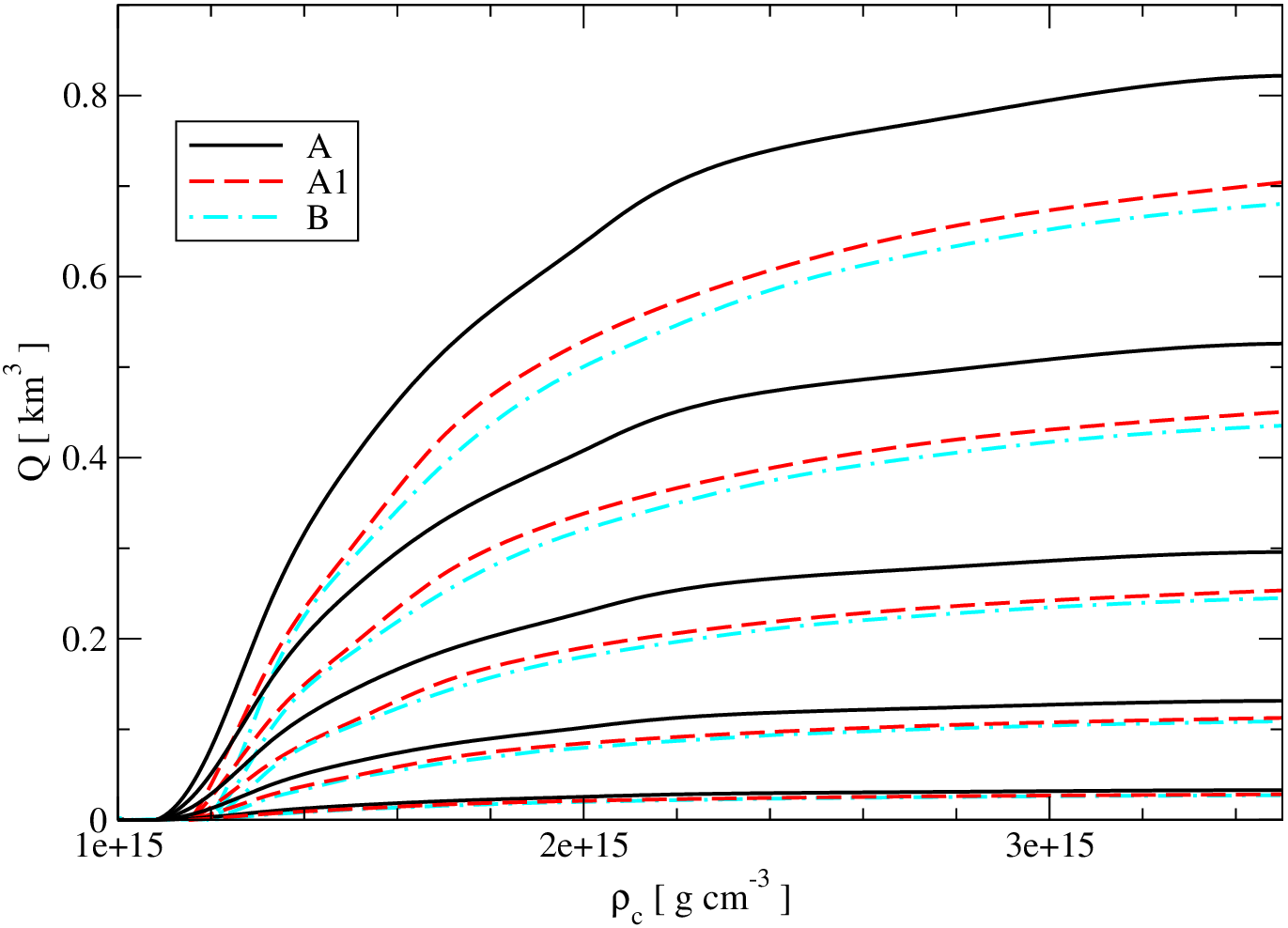}
\end{center}
\caption[]
{(Color online)
The quadrupole moments of stellar configurations computed 
according to Eqs. (\ref{eq:quadrupole}), (\ref{eq:mu_estimate}) 
and (\ref{eq:coefficients}) for  models  A, A1, and B as a function
of central density (the labeling of  models is as in Fig.~\ref{fig:n_P}). 
Each triple of curves
corresponds to the gap parameter value (from top to bottom) 
$\Delta =50$, 40, 30, 20, and 10 MeV. The magnitude of the 
breaking strain is $\bar\sigma_{\rm max}=10^{-2}$, and $Q_{\rm max}$ 
scales linearly with $\bar\sigma_{\rm max}$.
}
\label{fig:Q_dens_delta}
\end{figure}
\begin{figure}[tb]
\begin{center}
\includegraphics[height=8.0cm,width=\linewidth,angle=0]{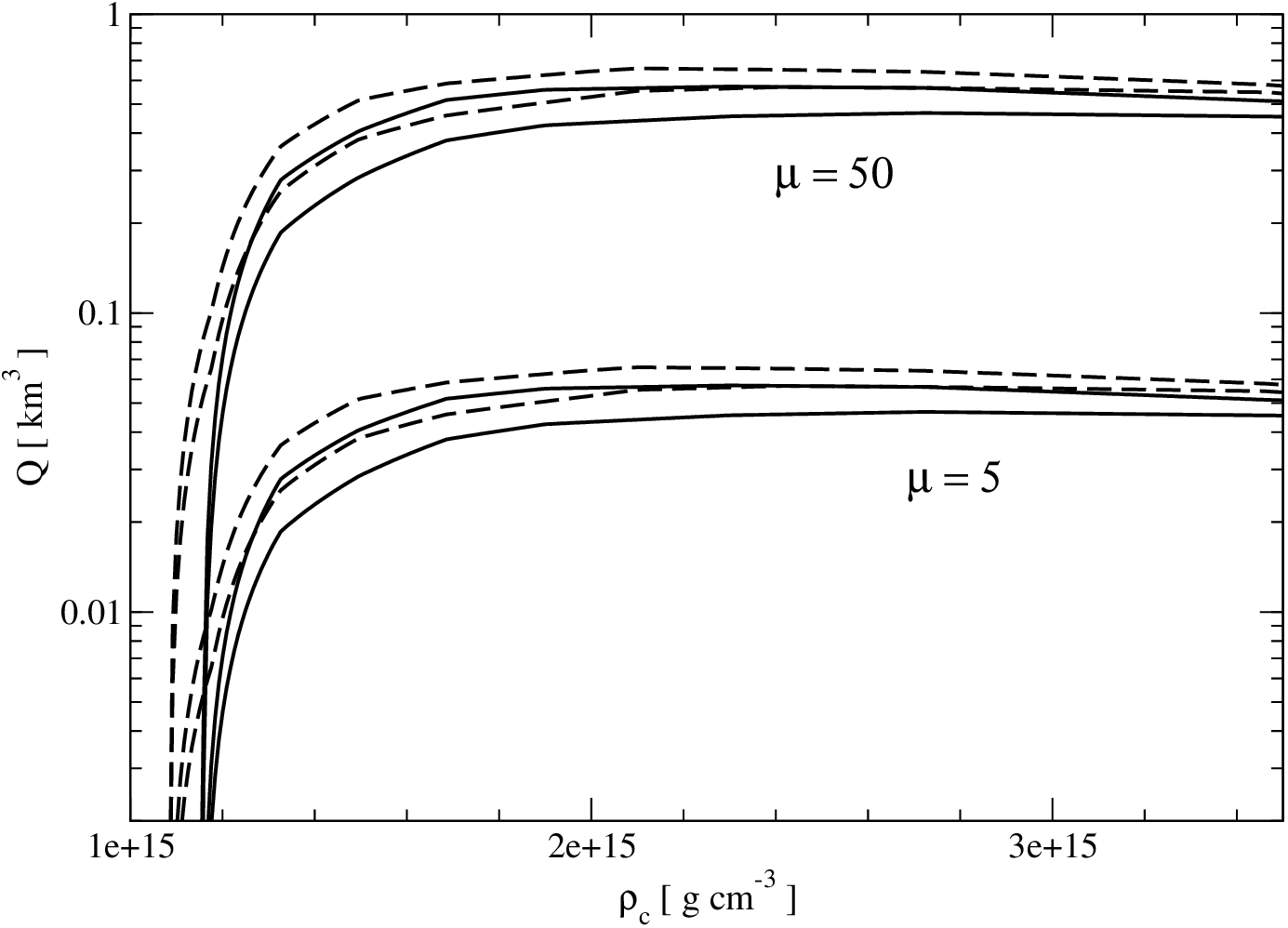}
\end{center}
\caption[]
{
The quadrupole moments of stellar configurations  (solid lines)
computed for fixed values of the shear modulus of CCS quark matter $\mu = 5$ 
and $\mu = 50$ MeV fm$^{-3}$ for model A as a function of central density.
The dashed lines represent the uniform-density incompressible matter 
approximation, Eq.~(\ref{eq:Q_approx}), where the masses and radii 
are set equal to that of the quark cores of microscopic models (the
values of the central density on the $x$ axis do not represent the 
central densities of these models).
}
\label{fig:Q_dens_comparison}
\end{figure}
Upon expanding the tensor $t_{ij}$ in spherical harmonics and integrating
(\ref{eq:perturbation}) with the boundary condition that assumes 
vanishing stress in the nuclear envelope surrounding the CCS quark core 
one finds [cf. Ref.~\cite{Ushomirsky:2000ax},  Eq. (64)]
\bea \label{eq:quadrupole}
Q_{22} &=& \int_0^{R_{\rm core}} \frac{dr r^3}{g(r)} \Bigg[
\frac{3}{2}(4-U) t_{rr}+\frac{1}{3}(6-U) t_{\Lambda}\nonumber\\
&+&\sqrt{\frac{3}{2}}\left(8-3U+\frac{1}{3}U^2-\frac{r}{3}\frac{dU}{dr}\right)
t_{r\perp}\Bigg],
\eea
where $U= 2+ d{\rm ln}g(r)/d{\rm ln} r$ and $t_{rr}$, $t_{\Lambda}$
and $t_{r\perp}$ are the coefficients of the expansion of the shear stress 
tensor in spherical harmonics \cite{Ushomirsky:2000ax}. 
These (unknown) coefficients can be 
determined if one assumes that the CCS core is maximally strained. 
The shear modulus is defined as $\mu = t_{ij}/2\sigma_{ij}$, where
$\sigma_{ij}$ is the strain tensor. The shear modulus of CCS matter 
has been estimated in Ref.~\cite{Mannarelli:2006fy} 
\be\label{eq:mu_estimate}
\mu = 2.47 ~{\rm MeV}~{\rm fm}^{-3}
\left(\frac{\Delta}{10~{\rm MeV}}\right)^2 
\left(\frac{\mu_q}{400~{\rm MeV}}\right)^2,
\ee
where $\Delta$ is the gap parameter, $\mu_q$ is the quark chemical 
potential. For a maximally strained CCS core 
\bea\label{eq:coefficients}
t_{rr}&=&2\mu \left(\frac{32\pi}{15}\right)^{1/2}\bar \sigma_{\max},\quad\quad
t_{r\perp}=2\mu \left(\frac{16\pi}{5}\right)^{1/2}\bar \sigma_{\max},\nonumber\\
t_{\Lambda}&=&2\mu \left(\frac{96\pi}{5}\right)^{1/2}\bar \sigma_{\max},
\eea
where $\bar\sigma_{\max}$ is the maximal value of the quantity 
$\bar\sigma^2 = \sigma_{ij}\sigma^{ij}/2$ (breaking strain). 
It is assumed that the strain in the CCS core is position 
independent. By combining Eq.~(\ref{eq:quadrupole}) and 
(\ref{eq:coefficients}) we obtain the maximal quadrupole moment 
$Q_{\rm max}$. If the core is  uniform and incompressible, then
\cite{Owen:05}
\be\label{eq:Q_approx}
\tilde Q_{\rm max} \simeq \frac{13 \mu \bar\sigma_{\max}R_{\rm core}^6}
{GM_{\rm core}}.
\ee
\begin{figure}[htb]
\begin{center}
\includegraphics[height=8.0cm,width=\linewidth,angle=0]{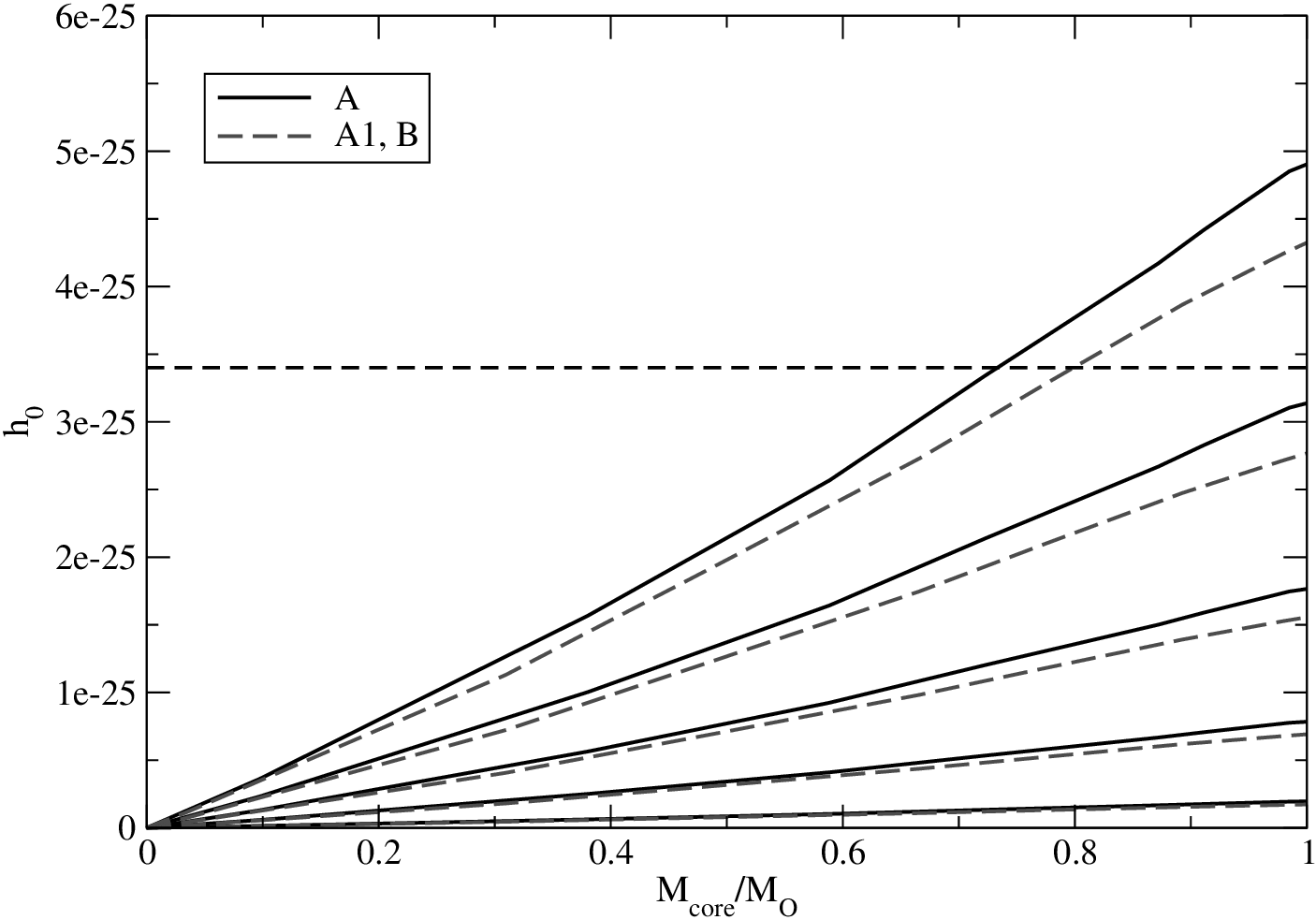}
\end{center}
\caption[]
{(Color online)
The strain of gravitational wave emission for a pulsar rotating 
at the Crab pulsar frequency $\nu = 29.6$ Hz at a distance of 2 kpc, 
computed according to Eqs. (\ref{eq:quadrupole}) and (\ref{eq:mu_estimate}) 
for models A and A1 (the results for models A1 and B overlap on the scale 
of this figure). Each triple of 
curves corresponds to the gap parameter value (from top to bottom) 
$\Delta =50$, 40, 30, 20, and 10 MeV. The magnitude of the breaking 
strain is $\bar\sigma_{\rm max}=10^{-4}$, and $h_0$ scales linearly 
with $\bar\sigma_{\rm max}$. The current upper limit on the strain 
of the Crab pulsar derived by the LIGO experiment is shown by the 
horizontal dashed line.
}
\label{fig:h_M_DELTA}
\end{figure}
\begin{figure}[bth]
\begin{center}
\includegraphics[height=8.0cm,width=\linewidth,angle=0]{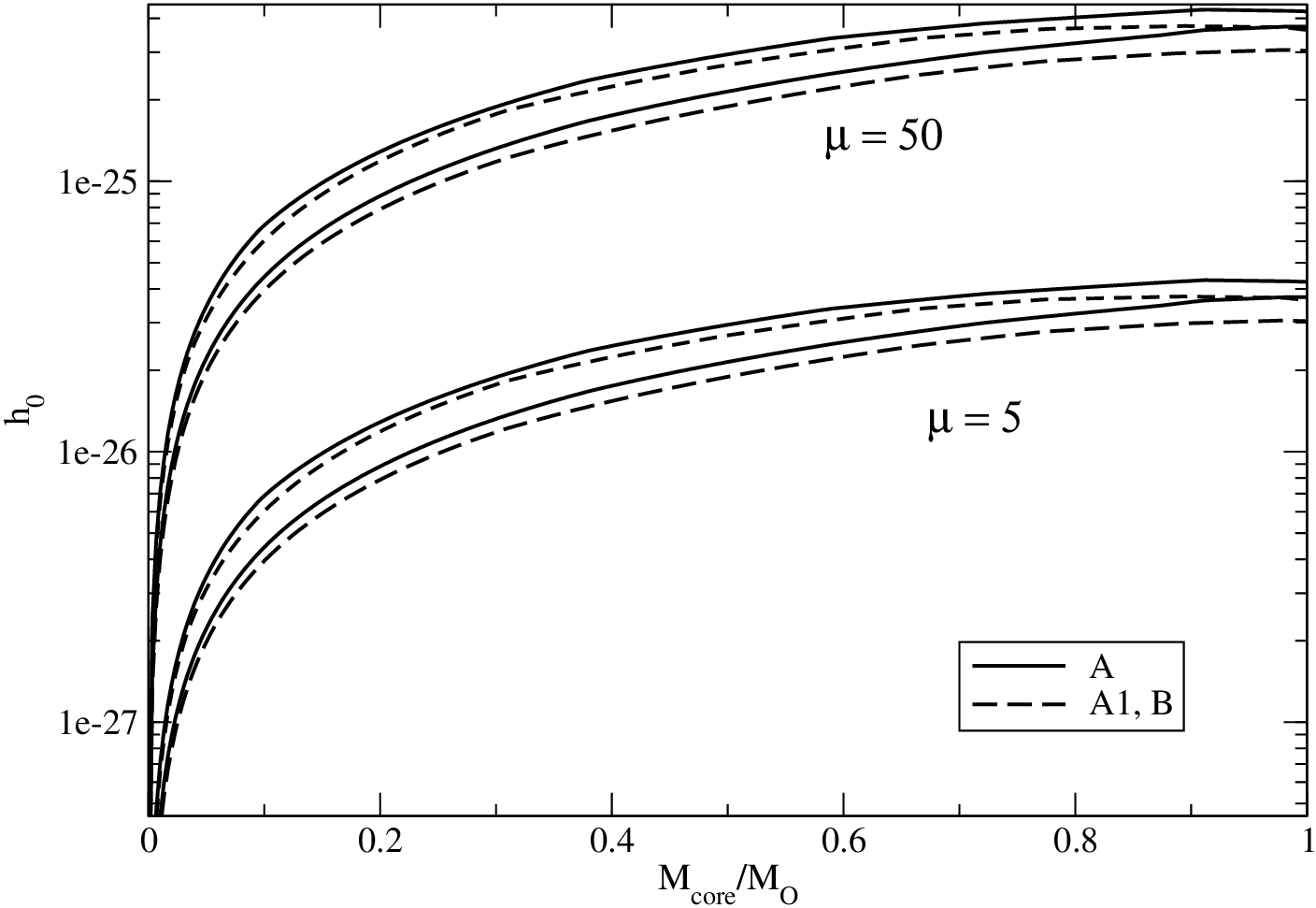}
\end{center}
\caption[]
{Same as in Fig.~\ref{fig:h_M_DELTA}, but for fixed  values of the 
shear modulus of CCS quark matter $\mu = 5$ and
$\mu = 50$ MeV fm$^{-3}$ for models A (solid lines) and A1 (dashed lines). 
The results for models A1 and B overlap on the scale of the figure.
For each pair of curves with the same value of $\mu$ and same equation 
of state the  lower curve is obtained
from the full expression for $Q_{\rm max}$, while the upper one  from 
the uniform-density incompressible fluid approximation,
Eq. (\ref{eq:Q_approx}). The latter approximation systematically 
overestimates the value of the quadrupole moment. 
The magnitude of the breaking strain is $\bar\sigma_{\rm max}=10^{-4}$;
 $h_0$ scales linearly with $\bar\sigma_{\rm max}$.
}
\label{fig:h_M_Vergleich}
\end{figure}

Figure~\ref{fig:Q_dens_delta} displays maximal quadrupole moments of a
sequences of hybrid stars derived by combining Eqs. (\ref{eq:quadrupole}),
(\ref{eq:mu_estimate}) and (\ref{eq:coefficients}). The quadrupole moment 
depends on the product of  the breaking strain and 
the shear modulus. The former will be treated as a free parameter in view
of the poor knowledge of this quantity; in Fig.~\ref{fig:Q_dens_delta} we
assume $\bar\sigma_{\rm max}=10^{-2}$. Since the equation of state fixes 
the quark chemical potential, the only parameter on which the shear modulus 
depends is the gap, which we will vary in the range $10\le \Delta \le
50$ MeV. It is assumed that the gap is independent of the density 
of the CCS quark phase. Treatments of the CCS phase that are based on the 
Ginzburg-Landau expansion near the critical temperature predict values 
of gaps that are in the 
range $5\le \Delta \le 25$ MeV. The low temperature CCS phase (more relevant 
for the problem at hand) is likely to support a larger, up to 100 MeV, gap.
The maximal quadrupole moment scales as $Q_{\rm max}\sim  \bar\sigma_{\rm max}
\Delta^2$ according to Eqs.~(\ref{eq:quadrupole})--(\ref{eq:coefficients}).

Figure \ref{fig:Q_dens_comparison} compares the quadrupole moments 
of quark cores of stellar configurations computed from the 
microscopic equations of state A, A1, and B with the quadrupole 
moments of uniform, incompressible objects computed via 
Eq.~(\ref{eq:Q_approx}). The masses and radii of the latter are 
set equal to that of the quark cores in the microscopic models. (Note that  
the central densities on the abscissa of Fig.~\ref{fig:Q_dens_comparison}
refer only to the microscopic models.) It is seen that the uniform, 
incompressible approximation systematically overestimates the magnitudes 
of the quadrupole moments, although numerically the error 
is not large. An accurate value of the quark core radius, that 
follows from our microscopic treatment, appears to be a more 
important factor in providing realistic values of 
$Q_{\rm max}$ because of $Q_{\rm max}\sim R_{\rm core}^6$ scaling.

Figure~\ref{fig:h_M_DELTA} displays the strain of gravitational wave 
emission given by Eq.~(\ref{eq:h0_Q22}) for models A, A1, and B as a
function of the core mass $M_{\rm core}$ (the results for
model A overlap with that of model B on the scale of the figure). 
Each pair of curves corresponds 
to the gap parameter value (from top to bottom)  $\Delta =50$, 40, 30, 
20, and 10 MeV. We assume rotation at the frequency of the
Crab pulsar $\nu = 29.6$ Hz at a distance of 2 kpc.
Since $h_0$ is a linear function of quadrupole moment $Q_{\rm max}$, 
its dependence on the gap function reflects  the quadratic dependence 
of $Q_{\rm max}$ on the gap function discussed
above. The magnitude of the breaking strain is $\bar\sigma_{\rm
max}=10^{-4}$.  The current upper limit on the characteristic strain of 
gravitational radiation from the Crab pulsar is 
3.4 $\times 10^{-25}$ and is shown in 
Fig.~\ref{fig:h_M_DELTA} by the horizontal dashed line. It is seen that 
the strain of gravitational wave emission is close to the upper limit 
for $\Delta = 50$ MeV.  If, however, the gaps are small 
more ``optimistic'' values of $10^{-3}\le \bar\sigma_{\rm max}
\le 10^{-2}$ will be needed to generate sizeable strain amplitude.
Clearly,  because of the large uncertainty in the actual 
value of $\bar\sigma_{\rm max}$ no definite conclusion can be drawn about 
the microscopic parameters of the CCS phase from 
the current upper limits on the 
strain of gravitational wave emission.
The current upper limit for gravitational waves from the Crab pulsar implies
$\bar\sigma_{\rm max}\Delta^2  \sim 0.25$ MeV$^2$ (under the assumptions 
of the present model).

Figure~\ref{fig:h_M_Vergleich} shows  the strain of gravitational 
wave emission $h_0$ for two approximations to the quadrupole deformations
(realistic vs incompressible), discussed above and illustrated
in Fig.~\ref{fig:Q_dens_comparison}. As in Fig.~\ref{fig:Q_dens_comparison},
the masses and radii of incompressible models are set equal to 
that of the quark cores derived from the microscopic models, i.e.,
the abscissa of Fig.~\ref{fig:h_M_Vergleich} refers to central density 
of the microscopic models only. It is seen that the uniform, 
incompressible approximation systematically overestimates the magnitude
of $h_0$ for all central densities.

\section{Conclusions}
\label{sec:conclusion}

We studied gravitational wave emission by hybrid compact (neutron) stars 
harboring maximally strained crystalline color-superconducting quark cores
on a basis of microscopic equations of state. In a first step, we
(re)constructed  hybrid configurations, which are composed of color 
superconducting quark matter at high and purely nuclear matter at low 
densities. The high-density quark matter is described in terms of the 
semimicroscopic Nambu--Jona-Lasinio 
model which includes pair correlations that lead 
to the three-flavor LOFF phase as the ground state of QCD at moderate 
densities. The low density nuclear phase is described in terms of hard 
relativistic equations of state based on the Dirac-Bruckner-Hartree-Fock 
theory. These configurations in static gravitational equilibrium arise 
as a ``second family'' of stable configurations. We find that the stable (with
respect to a collapse into a black hole) segment of the sequences have 
CCS quark cores with masses $0\le M_{\rm core} /M_{\odot}\le 0.75-0.88$ for 
central densities $1.3 \times 10^{15}\le \rho_c\le 2\times 10^{15}$
and radii up to 7 km. Consequently, about 1/3 of the mass of hybrid
configuration is concentrated in its quark core, which occupies about 
20 $\%$ of star's volume. 

In the second step, we considered the characteristic 
strain of gravitational wave emission due to deformations of the CCS 
quark core. If the quark core is maximally strained, the components of 
the stress tensor can be expressed through  product 
of the breaking strain and the shear modulus of the CCS quark 
matter. The quadrupole moment and the characteristic strain mainly depend 
on the not-well-known product $\bar\sigma_{\rm max}\Delta^2$.
The current upper limits imply that  $\bar\sigma_{\rm max}\Delta^2\le 0.25$
MeV$^2$, which can be accommodated by assuming either large gaps 
$\Delta \sim 50$ MeV and moderate values of $\bar\sigma_{\rm max}\sim 10^{-4}$ 
or small gaps $\sim 15$ MeV and more ``optimistic'' values of breaking strain, 
$\bar\sigma_{\rm max}\sim 10^{-3}$. In view of large 
uncertainty in the value of breaking strain $10^{-5} \le \bar\sigma_{\rm max}
\le 10^{-2}$ no definite conclusions can be made about the possible magnitude 
of the gap in the CCS phase.  However, if the gaps are large, $\Delta \ge 50$ MeV,
the values $\bar\sigma_{\rm max} \ge 10^{-3}$ are excluded by the present scenario 
of gravitational wave radiation. Note, however, that the assumption that the CCS 
core is maximally strained may not correspond to the actual situation in 
compact stars and the evolutionary avenues that may lead to such maximal 
deformations are not explored yet. One may only speculate that a rapidly 
spinning pulsar at birth may be triaxially stretched, a deformation that 
may be preserved within the star after the transition to the crystalline 
state. Nevertheless, we conclude that the characteristic strain amplitudes 
of gravitational waves produced by our models are well within the reach of 
the current gravitational wave detectors for a wide and reasonable range of 
parameters. 

\section*{Acknowledgements}
We are grateful to Mark Alford, Nils Andersson, Michael Buballa,
Nicola Ippolito, Krishna Rajagopal, Luciano Rezzolla, Dirk H. Rischke,  
Marco Ruggieri, Lars Samuelesson, J\"urgen Schaffner-Bielich,
Fridolin Weber and Graham Woan for useful interactions. 
This work was supported in part by the Deutsche Forschungsgemeinschaft.


\end{document}